\documentclass{article}
\usepackage{gdgspace, emlines2}
\advance\textheight by -\headsep\advance\textheight by -\headheight
\begin{document}
\large
\pagestyle{plain}
\onehalfspacing

\centerline{\LARGE Why the Afshar Experiment Does Not Refute Complementarity}\vskip 1.5cm
\centerline{R. E. Kastner}
\centerline{Department of Philosophy}
\centerline{University of Maryland}
\centerline{College Park, MD 20742}
\centerline{Version of April 26, 2005}
\vskip .5cm

ABSTRACT. A modified version of Young's experiment by Shahriar Afshar demonstrates that, 
prior to what appears to be a ``which-way'' measurement, an interference pattern exists. 
Afshar has claimed that this result constitutes a violation of the Principle of Complementarity. 
This paper discusses the implications of this experiment and considers how Cramer's 
Transactional Interpretation easily accomodates the result. It is also shown that 
the Afshar experiment is analogous in key respects to a spin one-half particle prepared 
as ``spin up along $\bf x$'', subjected to a nondestructive confirmation of that preparation, 
and post-selected in a specific state of spin along $\bf z$.
 The terminology ``which-way'' or ``which-slit'' is critiqued; it is argued that this
 usage by both Afshar and his critics is misleading and has contributed to confusion
 surrounding the interpretation of the experiment. Nevertheless, it is concluded that
 Bohr would have had no more problem accounting for the Afshar result than he would 
in accounting for the aforementioned pre- and post-selection spin experiment, 
in which the particle's preparation state is confirmed by a nondestructive measurement 
prior to post-selection. In addition, some new inferences about the interpretation
 of delayed choice experiments are drawn from the analysis.
\vskip 1cm
{\bf 1. Introduction.}
\vskip .5cm

The Young two-slit experiment is a famous illustration of
wave-particle duality: a quantum particle emitted toward a screen
with two small slits will produce an interference pattern on
a detecting screen downstream from the slits. On the other hand,
as has been repeatedly demonstrated, if one tries to obtain
``which-way'' or ``which slit'' information, the downstream
interference pattern vanishes and the distribution is the one
that would be expected for classical particles with no
wave aspect at all. A new photon two-slit experiment by Shahriar S. Afshar (2004) detects
an interference pattern in the region
between the slits and a final measurement characterized by
Afshar as revealing which slit the photon has gone through. Afshar interprets
his result as a falsification of Bohr's Principle of Complementarity (PC),
since it appears to give ``which-way'' information while
still detecting an interference pattern.

This paper analyzes the experiment and its resulting phenomena from
the perspective of the Transactional Interpretation (TI)
of John G. Cramer (1986), and it is argued that the TI picture provides
a natural way to understand the various phenomena involved.
It is argued that the Afshar experiment is analogous in key
respects to a standard spin-1/2 pre- and post-selection experiment,
and that Bohr would not have been at all shocked by the result.
\vskip .5cm
{\bf 2. First, a generic experiment}
\vskip .5cm
Before considering the details of Afshar's experiment and his
claims, let's consider the following generic two-state experiment.
Consider a Hilbert space spanned by two basis vectors we might call ``upper'' $|U\rangle$, and ``lower''
$|L\rangle$. Let's also define the ``superposition'' state $|S\rangle$, 
$$|S\rangle = {1\over\sqrt 2} [|U\rangle + |L\rangle] \eqno(1)$$

(For convenience, in the following we'll just use the capital letters without the
kets to designate a particular state).

Now, the experiment consists of preparing the state S at time $t_0$,
and then, at a later time $t_2$, measuring the observable whose eigenstates
are U and L. We don't doubt that, provided we have performed the usual
sharp measurement which clearly separates the states U and L (through an
appropriate interaction Hamiltonian coupling the quantum system and a macroscopic measuring
apparatus), that we have obtained
a legitimate result at $t_2$ for whether the system is in the state U or L,
at time $t_2$.

If this experiment sounds familiar, it's because it describes exactly what
goes on in a typical spin-measurement for a spin-1/2 particle with
respect to two orthogonal spatial directions, say $\bf x$ and $\bf z$. In terms
of the states corresponding to outcomes of 
``up/down along $\bf x$'' and ``up/down along $\bf z$,'' We just make the identification

$$|S\rangle  \equiv |x\uparrow\rangle$$

and

$$|U\rangle \equiv |z\uparrow\rangle, \quad |L\rangle \equiv |z\downarrow\rangle. \eqno (2 a,b,c)$$

Note that in this spin experiment, even though we allow that 
finding result ``U,'' or ``up along $\bf z$''
at $t_2$ is accurate in that it does describe the particle's spin state
at $t_2$, we {\it don't} typically use that result to assert that ``from the time of preparation
in state `S' (or `up along $\bf x$'), the particle really was `U' or
`up along $\bf z$'.'' If we do try
to make the above kind of retrodiction, we run into a paradox, because we can
also make a {\it pre}diction that the particle should ``really'' be ``up along $\bf x$'' during
the same interval of time.\footnote{\normalsize Such a prediction can
be obtained, for example, by using the Aharonov-Bergmann-Lebowitz (ABL) rule
for measurements performed on pre- and post-selected systems (ABL, 1964).} 
So we obtain two mathematically correct inferences
that such a particle is simultaneously in
eigenstates of different complementary observables. This paradox is described in a paper
by Albert, Aharonov and D'Amato (1985). It is also addressed effectively by
Cramer (1986), an analysis which we revisit in section 4.
\vskip .5cm
{\bf 3. The Afshar Experiment}
\vskip .5cm
The generic two-state experiment discussed above also describes what goes on
in the Afshar two-slit experiment without the wire grid (see Figure 1). The basis 
defined by the presence of the two slits can be
labeled by the vectors {$|U\rangle,|L\rangle$} as above. We can define
an observable corresponding to this basis, say $\cal O$. The lens serves to
provide for a sharp measurement of outcome either U or L at
$t_2$ (the time index corresponding to the placement of the
final screen at location $\sigma_2$). However---and this is crucial---since both slits
are open at the beginning of the experiment (at time $t_0$),
the photon is being prepared in the state S, corresponding to
a superposition of U and L; it is {\it not} in an eigenstate of
$\cal O$.\vskip 2cm

%TexCad Options
%\grade{\on}
%\emlines{\on}
%\beziermacro{\off}
%\reduce{\on}
%\snapping{\off}
%\quality{2.00}
%\graddiff{0.01}
%\snapasp{1}
%\zoom{1.00}
\special{em:linewidth 0.4pt}
\unitlength 1mm
\linethickness{0.4pt}
\begin{picture}(133.00,80.33)(20,40)
%\bezier{200}(80.33,123.33)(91.67,99.67)(80.33,78.33)
\emline{80.33}{123.33}{1}{81.42}{120.96}{2}
\emline{81.42}{120.96}{3}{82.38}{118.60}{4}
\emline{82.38}{118.60}{5}{83.24}{116.25}{6}
\emline{83.24}{116.25}{7}{83.97}{113.91}{8}
\emline{83.97}{113.91}{9}{84.60}{111.59}{10}
\emline{84.60}{111.59}{11}{85.11}{109.27}{12}
\emline{85.11}{109.27}{13}{85.50}{106.97}{14}
\emline{85.50}{106.97}{15}{85.78}{104.68}{16}
\emline{85.78}{104.68}{17}{85.95}{102.40}{18}
\emline{85.95}{102.40}{19}{86.00}{100.14}{20}
\emline{86.00}{100.14}{21}{85.94}{97.88}{22}
\emline{85.94}{97.88}{23}{85.76}{95.64}{24}
\emline{85.76}{95.64}{25}{85.47}{93.41}{26}
\emline{85.47}{93.41}{27}{85.06}{91.19}{28}
\emline{85.06}{91.19}{29}{84.54}{88.98}{30}
\emline{84.54}{88.98}{31}{83.91}{86.79}{32}
\emline{83.91}{86.79}{33}{83.16}{84.60}{34}
\emline{83.16}{84.60}{35}{82.29}{82.43}{36}
\emline{82.29}{82.43}{37}{81.31}{80.27}{38}
\emline{81.31}{80.27}{39}{80.33}{78.33}{40}
%\end
%\bezier{200}(80.33,123.33)(69.00,100.67)(80.67,79.00)
\emline{80.33}{123.33}{41}{79.25}{121.06}{42}
\emline{79.25}{121.06}{43}{78.29}{118.79}{44}
\emline{78.29}{118.79}{45}{77.44}{116.52}{46}
\emline{77.44}{116.52}{47}{76.71}{114.26}{48}
\emline{76.71}{114.26}{49}{76.09}{112.01}{50}
\emline{76.09}{112.01}{51}{75.59}{109.76}{52}
\emline{75.59}{109.76}{53}{75.21}{107.51}{54}
\emline{75.21}{107.51}{55}{74.94}{105.27}{56}
\emline{74.94}{105.27}{57}{74.79}{103.04}{58}
\emline{74.79}{103.04}{59}{74.75}{100.81}{60}
\emline{74.75}{100.81}{61}{74.83}{98.58}{62}
\emline{74.83}{98.58}{63}{75.03}{96.36}{64}
\emline{75.03}{96.36}{65}{75.34}{94.15}{66}
\emline{75.34}{94.15}{67}{75.77}{91.94}{68}
\emline{75.77}{91.94}{69}{76.32}{89.73}{70}
\emline{76.32}{89.73}{71}{76.98}{87.53}{72}
\emline{76.98}{87.53}{73}{77.75}{85.34}{74}
\emline{77.75}{85.34}{75}{78.65}{83.15}{76}
\emline{78.65}{83.15}{77}{79.66}{80.96}{78}
\emline{79.66}{80.96}{79}{80.67}{79.00}{80}
%\end
\emline{34.33}{122.67}{81}{34.33}{111.33}{82}
\emline{34.33}{105.33}{83}{34.33}{94.00}{84}
\emline{34.33}{89.00}{85}{34.33}{77.67}{86}
\emline{19.00}{115.67}{87}{19.00}{83.33}{88}
\emline{16.00}{115.33}{89}{16.00}{83.00}{90}
%\vector(16.00,98.67)(24.00,98.67)
\put(24.00,98.67){\vector(1,0){0.2}}
\emline{16.00}{98.67}{91}{24.00}{98.67}{92}
%\end
\emline{129.33}{123.00}{93}{129.33}{77.33}{94}
\emline{35.33}{110.33}{95}{80.67}{110.33}{96}
\emline{80.67}{110.33}{97}{129.33}{91.33}{98}
\emline{35.00}{107.33}{99}{80.67}{90.33}{100}
\emline{80.67}{90.33}{101}{129.33}{90.33}{102}
\emline{34.67}{90.33}{103}{80.33}{90.33}{104}
\emline{80.33}{90.33}{105}{129.00}{110.67}{106}
\emline{34.67}{92.67}{107}{80.67}{110.33}{108}
\emline{80.67}{110.33}{109}{129.33}{110.33}{110}
\put(71.00,85.00){\dashbox{2.00}(1.00,31.00)[cc]{ }}
\put(71.67,126.67){\makebox(0,0)[cc]{$\sigma_1$}}
\put(129.33,127.33){\makebox(0,0)[cc]{$\sigma_2$}}
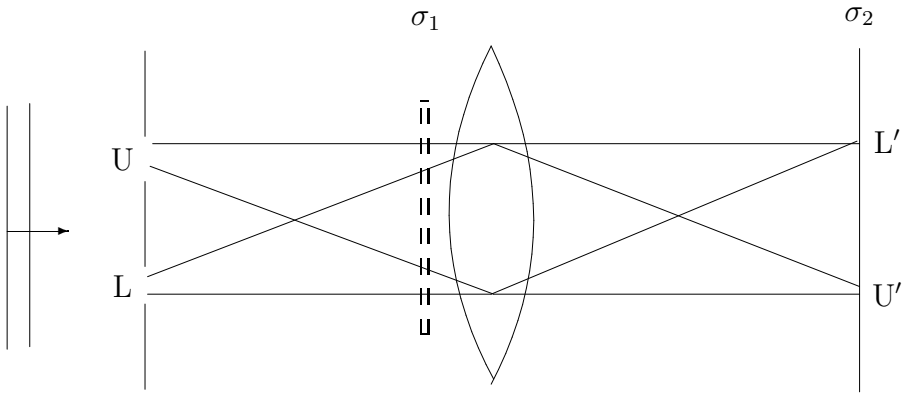
\put(76.67,55.00){\makebox(0,0)[cc]{Figure 1. The setup for the Afshar experiment.}}
\put(31.33,108.33){\makebox(0,0)[cc]{U}}
\put(31.33,91.33){\makebox(0,0)[cc]{L}}
\put(133.00,110.67){\makebox(0,0)[cc]{L$^\prime$}}
\put(133.00,90.33){\makebox(0,0)[cc]{U$^\prime$}}
\end{picture}

Afshar places a thin wire grid just before the lens, 
intercepting only those regions where a minima would occur
in the calculated interference pattern at $\sigma_1$. 
When the wire grid is in place, corresponding
to time $t_1$ (between $t_0$ and $t_2$), it performs a nondestructive
confirmation measurement of the prepared state S (taking
into account the unitary evolution of that state to the
spatial distribution at $\sigma_1$).\footnote{\normalsize Because if the
photon were not in state S, but instead U or L, the wires would block
photons expected to be present at those locations.} This is exactly analogous,
in our spin-measurement experiment, to inserting an additional x-oriented
Stern-Gerlach device at time $t_1$, and confirming our preparation
of the particle in the state $|x\uparrow\rangle$. But no one would deny
that we can still make a subsequent z-spin measurement at $t_2$ and
count the result (either U or L) as giving information about the
particle's spin along z at that time.

Similarly, in the Afshar experiment, a photon is prepared at time $t_0$ in
the state S (analogous to ``up along $\bf x$''). When the wire grid is
in place at $\sigma_1$ $(t_1)$, it confirms the preparation of the
photon in that state. Finally, at $\sigma_2$ $(t_2)$, the photon
is post-selected via a measurement of $\cal O$ in either state U or L. There is nothing
illegitimate about that measurement, any more than there is something
illegitimate about measuring the observable $J_z$---that is, 
the spin of a particle along $\bf z$---at $t_2$ after we
have prepared it as ``up along $\bf x$'' at $t_0$ and reconfirmed it
as still ``up along $\bf x$'' at $t_1$.

Nevertheless, we tend to want to reject the idea that Afshar's final measurement
really is a measurement of the observable $\cal O$ corresponding to
the slit basis \{U,L\}. Why? Because we are used to thinking 
of that measurement as a ``which-way'' measurement allowing us to infer
which slit the photon ``really'' went through. But, in view of
the interference pattern detected by the grid, we know that
would be nonsense---the photon was clearly in a superposition of
slits at $t_1$, so it couldn't have gone through only one or the
other slit.

Nevertheless, even with the grid removed, since the photon is prepared
in a superposition S, the measurement at the final screen at $t_2$
{\it never really is} a ``which-way'' measurement (the term traditionally
attached to the slit-basis observable $\cal O$), 
because it cannot tell us ``which slit the photon really went through.''
That is, {\it a good measurement of the slit basis observable is not equivalent
to the disclosure of a definite trajectory through a particular slit, which
is what the terminology ``which-way'' suggests.}
Recall that, in the spin case, we should {\it not} conclude that a particle
prepared as ``up along $\bf x$'' and post-selected in the state ``up along $\bf z$''
was always ``really'' up along $\bf z$ throughout the time interval between
$t_0$ and $t_2$ (because
we could just as easily {\it predict} that it should be ``up along $\bf x$''
throughout that same interval). 

So we have a double standard: on the one hand, we recognize that we
should be cautious about retrodicting the final spin result
along $\bf z$ to a time just after it was in an eigenstate of the complementary
observable ``spin along $\bf x$,'' but we still want to think of
a photon's trajectory with respect to a particular
slit as ``retroactively'' determinate (in the Afshar setup without
the wire grid) when it was in fact prepared
in a superposition of slit locations. This double standard is revealed by our use of
the term ``which-way'' when what we are really referring to is simply a measurement of
the slit basis observable.

A specific argument commonly advanced against Afshar's final measurement
being a legitimate slit basis observable measurement is the following: 
``if only one slit were open, the wire grid would scatter photons into
the wrong part of the final screen.''\footnote{This is essentially the
argument made by Unruh (2004).} This is certainly true.
But notice that it is exactly analogous to the following statement:  

Statement A: ``{\it if} a spin-1/2 particle were prepared at $t_0$ in the state ``up along $\bf z$,''
then a measurement of spin along $\bf x$ made at $t_1$, and finally a measurement
of spin along $\bf z$ were made at $t_2$, then the particle might not
be reconfirmed in the original state ``up along $\bf z$''. 

This is of course true, but it doesn't mean that
that final measurement was not a legitimate z-spin measurement.
It just means that we can't assert that the particle was
always ``up (or down) along $\bf z$'' throughout the experiment.
But the point is that we shouldn't be doing that anyway in
the Afshar setup even without the wire grid at $t_1$---since the photon
starts out in the state S, not the state U or D, and thus was never in a determinate
state with respect to the slit observable (until $t_2$).

Nevertheless, if Afshar is taken as claiming that the final measurement tells which way the photon
``really'' went, that claim is of course false. Yet Afshar's final measurement does
qualify as just as good a measurement of the slit-basis observable $\cal O$ as does the final
spin measurement of $J_z$ in the analogous spin case. For, when we preselect
the particle as ``up along x'', confirm it in the same state by
another x-spin measurement, and make a subsequent final measurement
of $J_z$ (a procedure formally identical to what goes
on in the Afshar experiment), we still consider the latter a perfectly good $J_z$ measurement.
That is, nobody presents Statement A as an objection to the final
measurement's being considered a legitimate measurement of a particle's spin along $\bf z$.

Thus, the traditional term ``which-way measurement'' as employed both by Afshar and
many of his critics is misleading: a post-selection based on a perfectly good ``slit basis'' observable
measurement {\it never} gives us
license to attribute a localizable ``which-slit'' trajectory to a particle
actually going through {\it both} slits at $t_0$, whether or not
the grid is in place at $\sigma_1$. It just characterizes a situation in which a photon, prepared
in a superposition of slit locations, is forced to drop out of that superposition
and pick a final spot to land in (equivalently, to pick the state U or L). 
 
These considerations will hopefully become more clear when
Afshar's experiment is analyzed from the point of view of
Cramer's Transactional Interpretation, in the next section.
In any case, from the above argument, clearly Bohr would have
no problem with the phenomena reported in the Afshar experiment,
any more than he would have a problem with the analogous spin
experiment. If the photon in question demonstrates ``complementary wave
and particle aspects in the same experiment,'' so does our spin one-half
particle demonstrate ``complementary spin-x and spin-z behavior
in the same experiment.''

\vskip .8cm
{\bf 4. Afshar's experiment under the Transactional Interpretation}
\vskip .2cm
Cramer's TI is presented in a comprehensive manner in his (1986).
Under TI, a source emits an ``offer wave'' which is equivalent
to the usual Schr\"odinger wave; but when that offer wave
is absorbed, the absorber sends a time-reversed ``confirmation wave''
back in time to the source. Any observable events occur as
a result of a ``transaction'' between the two types of
waves. In any given experiment, there may be many possible
transactions, but in general only one can be realized.

Let us consider Figure 14 in Cramer's (1986), reproduced here as 
Figure 2.\footnote{\normalsize The notation has been slightly
altered to correspond to the discussion in this paper.}
This figure is used in conjunction with Cramer's discussion
of the pre- and post-selection puzzle presented by Aharonov,
Albert, and D'Amato (1985), and mentioned in section 2 above.
In this paradox, a polarized photon or spin-1/2 particle, i.e., a two-state system, seems
to have a probability of unity for being in either
the preselection state, for example, ``horizontally polarized''
($|H\rangle$) or the post-selection state, ``right circularly polarized'' ($|R\rangle$),
which appears to violate the uncertainty principle. \vskip 2cm

%TexCad Options
%\grade{\on}
%\emlines{\on}
%\beziermacro{\off}
%\reduce{\on}
%\snapping{\off}
%\quality{2.00}
%\graddiff{0.01}
%\snapasp{1}
%\zoom{1.00}
\special{em:linewidth 0.4pt}
\unitlength 1.00mm
\linethickness{0.4pt}
\begin{picture}(102.00,75.33)(0,50)
\emline{20.00}{127.67}{1}{20.00}{55.33}{2}
\emline{41.00}{127.67}{3}{41.00}{55.33}{4}
\emline{81.00}{128.00}{5}{81.00}{55.67}{6}
\emline{102.00}{128.00}{7}{102.00}{55.67}{8}
%\vector(24.33,117.33)(97.33,117.33)
\put(97.33,117.33){\vector(1,0){0.2}}
\emline{24.33}{117.33}{9}{97.33}{117.33}{10}
%\end
%\vector(97.33,111.33)(24.33,111.33)
\put(24.33,111.33){\vector(-1,0){0.2}}
\emline{97.33}{111.33}{11}{24.33}{111.33}{12}
%\end
%\vector(24.67,93.67)(97.67,93.67)
\put(97.67,93.67){\vector(1,0){0.2}}
\emline{24.67}{93.67}{13}{97.67}{93.67}{14}
%\end
%\vector(97.67,87.67)(24.67,87.67)
\put(24.67,87.67){\vector(-1,0){0.2}}
\emline{97.67}{87.67}{15}{24.67}{87.67}{16}
%\end
%\vector(25.00,69.33)(98.00,69.33)
\put(98.00,69.33){\vector(1,0){0.2}}
\emline{25.00}{69.33}{17}{98.00}{69.33}{18}
%\end
%\vector(98.00,63.33)(25.00,63.33)
\put(25.00,63.33){\vector(-1,0){0.2}}
\emline{98.00}{63.33}{19}{25.00}{63.33}{20}
%\end
\put(20.00,132.00){\makebox(0,0)[cc]{{\bf E}}}
\put(41.00,132.00){\makebox(0,0)[cc]{{\bf H}}}
\put(81.00,132.33){\makebox(0,0)[cc]{{\bf R}}}
\put(102.00,132.33){\makebox(0,0)[cc]{{\bf D}}}
\put(12.00,114.00){\makebox(0,0)[cc]{(a)}}
\put(11.67,90.33){\makebox(0,0)[cc]{(b)}}
\put(11.67,66.00){\makebox(0,0)[cc]{(c)}}
\emline{60.33}{97.00}{21}{60.33}{84.33}{22}
\emline{60.33}{73.00}{23}{60.33}{60.33}{24}
\put(60.33,100.00){\makebox(0,0)[cc]{{\bf H}}}
\put(60.67,75.67){\makebox(0,0)[cc]{{\bf R}}}
\put(31.67,121.33){\makebox(0,0)[cc]{$|SV\rangle$}}
\put(32.00,96.33){\makebox(0,0)[cc]{$|SV\rangle$}}
\put(32.00,72.33){\makebox(0,0)[cc]{$|SV\rangle$}}
\put(60.00,121.33){\makebox(0,0)[cc]{$|H\rangle$}}
\put(51.00,96.67){\makebox(0,0)[cc]{$|H\rangle$}}
\put(71.33,96.67){\makebox(0,0)[cc]{$|H\rangle$}}
\put(51.00,72.00){\makebox(0,0)[cc]{$|H\rangle$}}
\put(89.67,121.67){\makebox(0,0)[cc]{${1\over\sqrt 2}|R\rangle$}}
\put(89.67,97.33){\makebox(0,0)[cc]{${1\over\sqrt 2}|R\rangle$}}
\put(89.67,73.00){\makebox(0,0)[cc]{${1\over\sqrt 2}|R\rangle$}}
\put(71.33,72.33){\makebox(0,0)[cc]{${1\over\sqrt 2}|R\rangle$}}
\put(89.33,107.33){\makebox(0,0)[cc]{${1\over\sqrt 2}\langle R|$}}
\put(60.00,107.67){\makebox(0,0)[cc]{${1\over\sqrt 2}\langle R|$}}
\put(89.33,83.67){\makebox(0,0)[cc]{${1\over\sqrt 2}\langle R|$}}
\put(89.33,60.00){\makebox(0,0)[cc]{${1\over\sqrt 2}\langle R|$}}
\put(71.00,83.00){\makebox(0,0)[cc]{${1\over\sqrt 2}\langle R|$}}
\put(71.33,59.67){\makebox(0,0)[cc]{${1\over\sqrt 2}\langle R|$}}
\put(32.00,107.67){\makebox(0,0)[cc]{${1\over 2}\langle H|$}}
\put(32.00,84.00){\makebox(0,0)[cc]{${1\over 2}\langle H|$}}
\put(32.00,59.33){\makebox(0,0)[cc]{${1\over 2}\langle H|$}}
\put(51.00,83.00){\makebox(0,0)[cc]{${1\over 2}\langle H|$}}
\put(51.67,59.33){\makebox(0,0)[cc]{${1\over\sqrt 2}\langle R|$}}
%\circle(51.00,90.00){20.06}
\emline{51.00}{100.03}{25}{53.16}{99.79}{26}
\emline{53.16}{99.79}{27}{55.21}{99.10}{28}
\emline{55.21}{99.10}{29}{57.07}{97.98}{30}
\emline{57.07}{97.98}{31}{58.64}{96.49}{32}
\emline{58.64}{96.49}{33}{59.86}{94.70}{34}
\emline{59.86}{94.70}{35}{60.67}{92.69}{36}
\emline{60.67}{92.69}{37}{61.02}{90.55}{38}
\emline{61.02}{90.55}{39}{60.90}{88.38}{40}
\emline{60.90}{88.38}{41}{60.32}{86.29}{42}
\emline{60.32}{86.29}{43}{59.31}{84.38}{44}
\emline{59.31}{84.38}{45}{57.91}{82.72}{46}
\emline{57.91}{82.72}{47}{56.19}{81.41}{48}
\emline{56.19}{81.41}{49}{54.22}{80.50}{50}
\emline{54.22}{80.50}{51}{52.10}{80.03}{52}
\emline{52.10}{80.03}{53}{49.94}{80.02}{54}
\emline{49.94}{80.02}{55}{47.82}{80.49}{56}
\emline{47.82}{80.49}{57}{45.85}{81.39}{58}
\emline{45.85}{81.39}{59}{44.12}{82.70}{60}
\emline{44.12}{82.70}{61}{42.71}{84.35}{62}
\emline{42.71}{84.35}{63}{41.69}{86.26}{64}
\emline{41.69}{86.26}{65}{41.11}{88.35}{66}
\emline{41.11}{88.35}{67}{40.99}{90.51}{68}
\emline{40.99}{90.51}{69}{41.33}{92.66}{70}
\emline{41.33}{92.66}{71}{42.13}{94.67}{72}
\emline{42.13}{94.67}{73}{43.34}{96.47}{74}
\emline{43.34}{96.47}{75}{44.91}{97.96}{76}
\emline{44.91}{97.96}{77}{46.76}{99.09}{78}
\emline{46.76}{99.09}{79}{48.82}{99.79}{80}
\emline{48.82}{99.79}{81}{51.00}{100.03}{82}
%\end
%\put(60.67,42.33){\makebox(0,0)[cc]
%{Figure 2. Cramer's diagram of a photon/polarizer experiment.}}
%\circle(71.00,66.67){20.35}
\emline{71.00}{76.84}{83}{73.18}{76.61}{84}
\emline{73.18}{76.61}{85}{75.26}{75.91}{86}
\emline{75.26}{75.91}{87}{77.14}{74.78}{88}
\emline{77.14}{74.78}{89}{78.74}{73.27}{90}
\emline{78.74}{73.27}{91}{79.98}{71.46}{92}
\emline{79.98}{71.46}{93}{80.79}{69.43}{94}
\emline{80.79}{69.43}{95}{81.16}{67.26}{96}
\emline{81.16}{67.26}{97}{81.05}{65.07}{98}
\emline{81.05}{65.07}{99}{80.48}{62.96}{100}
\emline{80.48}{62.96}{101}{79.46}{61.01}{102}
\emline{79.46}{61.01}{103}{78.05}{59.33}{104}
\emline{78.05}{59.33}{105}{76.31}{57.99}{106}
\emline{76.31}{57.99}{107}{74.33}{57.05}{108}
\emline{74.33}{57.05}{109}{72.19}{56.56}{110}
\emline{72.19}{56.56}{111}{70.00}{56.54}{112}
\emline{70.00}{56.54}{113}{67.85}{56.99}{114}
\emline{67.85}{56.99}{115}{65.85}{57.89}{116}
\emline{65.85}{57.89}{117}{64.09}{59.20}{118}
\emline{64.09}{59.20}{119}{62.65}{60.85}{120}
\emline{62.65}{60.85}{121}{61.60}{62.78}{122}
\emline{61.60}{62.78}{123}{60.98}{64.88}{124}
\emline{60.98}{64.88}{125}{60.83}{67.07}{126}
\emline{60.83}{67.07}{127}{61.15}{69.24}{128}
\emline{61.15}{69.24}{129}{61.94}{71.29}{130}
\emline{61.94}{71.29}{131}{63.14}{73.13}{132}
\emline{63.14}{73.13}{133}{64.70}{74.66}{134}
\emline{64.70}{74.66}{135}{66.56}{75.83}{136}
\emline{66.56}{75.83}{137}{68.63}{76.56}{138}
\emline{68.63}{76.56}{139}{71.00}{76.84}{140}
%\end
\end{picture}

\singlespacing \normalsize
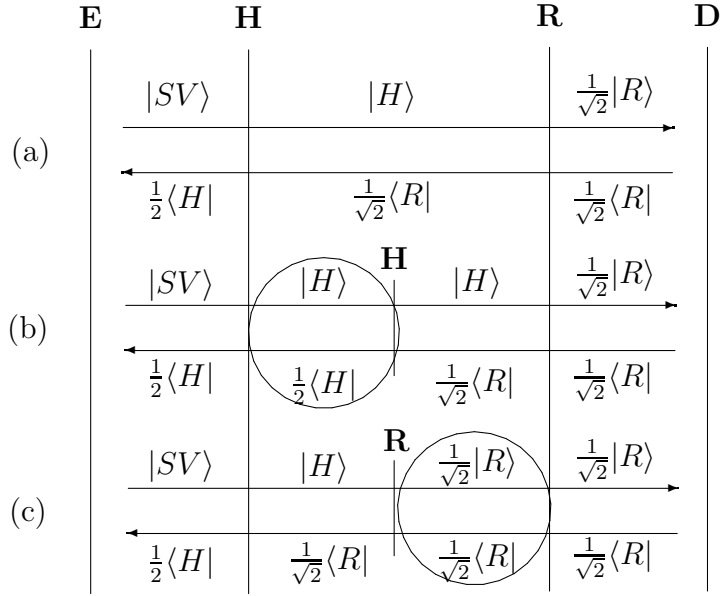
\noindent Figure 2. Cramer's diagram of a pre- and post-selection experiment
with polarized light.
In (a), no intervening measurement is performed;
in (b), an intervening measurement of horizontal
linear
polarization (H) is performed; and in (c)
a measurement of right-circular polarization (R) is performed.
\vskip .5cm\onehalfspacing\large

In the figure, the symbol $|SV\rangle$
refers to the initial unpolarized beam emanating from the 
source\footnote{\normalsize We neglect the factor of 1/2 arising from
the attenuation of the unpolarized beam when preselected.}
; in the Afshar case it would refer to the photon's
state prior to entering the slits. The four vertical lines
labeled E, H, R, and D represent the four times when
the photon leaves the emitter E, is selected in state $|H\rangle$,
is post-selected in state $|R\rangle$, and is absorbed by
detector D. 

Cramer points
out, via his figure, that it is important to consider
whether an intervening measurement is actually made,
since under TI the nature of the offer and confirmation waves will
differ depending on the situation (see caption).  
In particular, claims about the possession of definite properties
can only be made when the offer and confirmation waves
agree. For example, in 2(b), under TI the photon is determinately
in a state of horizontal polarization only in the interval
between preselection and the intervening measurement of H.
(The circles indicate where the photon's state is considered
determinate.)

In Figure 3 we have adapted Cramer's Figure 14
to reflect the Afshar setup, which is analogous
in key respects. In keeping with the discussion in
section 2, let us label the two slits U and L, with corresponding
states $|U\rangle$, etc.
Figure 3(a) schematically represents the essence of the Afshar
experiment without the wire grid at $\sigma_1$ if we label the initial 
or prepared state $|S\rangle = {1\over\sqrt 2} (|U\rangle + |L\rangle)$
 of the Afshar photon after it enters the two slits and the 
final (post-selected) state to be, say, $|U\rangle$,
after the which-way measurement is made and the photon lands
at U$^\prime$. (Technically, the post-selection
in U and the final detection D should be at the same location to conform
to the Afshar setup.) As the figure shows,
according to TI there is an offer wave $|S\rangle$ directed toward the
future in the usual way, and a confirmation wave $\langle U|$ directed
toward the past. Cramer points out in his Fig. 14(a) that the photon's
intermediate ontological state is ambiguous because the offer and confirmation
waves are different throughout the interval; the same applies
to the Afshar setup with no grid.\vskip 2cm

%TexCad Options
%\grade{\on}
%\emlines{\on}
%\beziermacro{\off}
%\reduce{\on}
%\snapping{\off}
%\quality{2.00}
%\graddiff{0.01}
%\snapasp{1}
%\zoom{1.00}
\special{em:linewidth 0.4pt}
\unitlength 1.00mm
\linethickness{0.4pt}
\begin{picture}(102.00,65.33)(0,60)
\emline{20.00}{127.67}{1}{20.00}{55.33}{2}
\emline{41.00}{127.67}{3}{41.00}{55.33}{4}
\emline{81.00}{128.00}{5}{81.00}{55.67}{6}
\emline{102.00}{128.00}{7}{102.00}{55.67}{8}
%\vector(24.33,117.33)(97.33,117.33)
\put(97.33,117.33){\vector(1,0){0.2}}
\emline{24.33}{117.33}{9}{97.33}{117.33}{10}
%\end
%\vector(97.33,111.33)(24.33,111.33)
\put(24.33,111.33){\vector(-1,0){0.2}}
\emline{97.33}{111.33}{11}{24.33}{111.33}{12}
%\end
%\vector(24.67,93.67)(97.67,93.67)
\put(97.67,93.67){\vector(1,0){0.2}}
\emline{24.67}{93.67}{13}{97.67}{93.67}{14}
%\end
%\vector(97.67,87.67)(24.67,87.67)
\put(24.67,87.67){\vector(-1,0){0.2}}
\emline{97.67}{87.67}{15}{24.67}{87.67}{16}
%\end
%\vector(25.00,69.33)(98.00,69.33)
\put(98.00,69.33){\vector(1,0){0.2}}
\emline{25.00}{69.33}{17}{98.00}{69.33}{18}
%\end
%\vector(98.00,63.33)(25.00,63.33)
\put(25.00,63.33){\vector(-1,0){0.2}}
\emline{98.00}{63.33}{19}{25.00}{63.33}{20}
%\end
\put(20.00,132.00){\makebox(0,0)[cc]{{\bf E}}}
\put(41.00,132.00){\makebox(0,0)[cc]{{\bf S}}}
\put(81.00,132.33){\makebox(0,0)[cc]{{\bf U}}}
\put(102.00,132.33){\makebox(0,0)[cc]{{\bf D}}}
\put(12.00,114.00){\makebox(0,0)[cc]{(a)}}
\put(11.67,90.33){\makebox(0,0)[cc]{(b)}}
\put(11.67,66.00){\makebox(0,0)[cc]{(c)}}
\emline{60.33}{97.00}{21}{60.33}{84.33}{22}
\emline{60.33}{73.00}{23}{60.33}{60.33}{24}
\put(60.33,100.00){\makebox(0,0)[cc]{{\bf S}}}
\put(60.67,75.67){\makebox(0,0)[cc]{{\bf U}}}
\put(31.67,121.33){\makebox(0,0)[cc]{$|SV\rangle$}}
\put(32.00,96.33){\makebox(0,0)[cc]{$|SV\rangle$}}
\put(32.00,72.33){\makebox(0,0)[cc]{$|SV\rangle$}}
\put(60.00,121.33){\makebox(0,0)[cc]{$|S\rangle$}}
\put(51.00,96.67){\makebox(0,0)[cc]{$|S\rangle$}}
\put(71.33,96.67){\makebox(0,0)[cc]{$|S\rangle$}}
\put(51.00,72.00){\makebox(0,0)[cc]{$|S\rangle$}}
\put(89.67,121.67){\makebox(0,0)[cc]{${1\over\sqrt 2}|U\rangle$}}
\put(89.67,97.33){\makebox(0,0)[cc]{${1\over\sqrt 2}|U\rangle$}}
\put(89.67,73.00){\makebox(0,0)[cc]{${1\over\sqrt 2}|U\rangle$}}
\put(71.33,72.33){\makebox(0,0)[cc]{${1\over\sqrt 2}|U\rangle$}}
\put(89.33,107.33){\makebox(0,0)[cc]{${1\over\sqrt 2}\langle U|$}}
\put(60.00,107.67){\makebox(0,0)[cc]{${1\over\sqrt 2}\langle U|$}}
\put(89.33,83.67){\makebox(0,0)[cc]{${1\over\sqrt 2}\langle U|$}}
\put(89.33,60.00){\makebox(0,0)[cc]{${1\over\sqrt 2}\langle U|$}}
\put(71.00,83.00){\makebox(0,0)[cc]{${1\over\sqrt 2}\langle U|$}}
\put(71.33,59.67){\makebox(0,0)[cc]{${1\over\sqrt 2}\langle U|$}}
\put(32.00,107.67){\makebox(0,0)[cc]{${1\over 2}\langle S|$}}
\put(32.00,84.00){\makebox(0,0)[cc]{${1\over 2}\langle S|$}}
\put(32.00,59.33){\makebox(0,0)[cc]{${1\over 2}\langle S|$}}
\put(51.00,83.00){\makebox(0,0)[cc]{${1\over 2}\langle S|$}}
\put(51.67,59.33){\makebox(0,0)[cc]{${1\over\sqrt 2}\langle U|$}}
%\circle(51.00,90.00){20.06}
\emline{51.00}{100.03}{25}{53.16}{99.80}{26}
\emline{53.16}{99.80}{27}{55.21}{99.10}{28}
\emline{55.21}{99.10}{29}{57.07}{97.99}{30}
\emline{57.07}{97.99}{31}{58.64}{96.50}{32}
\emline{58.64}{96.50}{33}{59.86}{94.70}{34}
\emline{59.86}{94.70}{35}{60.66}{92.69}{36}
\emline{60.66}{92.69}{37}{61.01}{90.55}{38}
\emline{61.01}{90.55}{39}{60.90}{88.39}{40}
\emline{60.90}{88.39}{41}{60.32}{86.30}{42}
\emline{60.32}{86.30}{43}{59.31}{84.38}{44}
\emline{59.31}{84.38}{45}{57.91}{82.73}{46}
\emline{57.91}{82.73}{47}{56.18}{81.41}{48}
\emline{56.18}{81.41}{49}{54.22}{80.50}{50}
\emline{54.22}{80.50}{51}{52.10}{80.03}{52}
\emline{52.10}{80.03}{53}{49.93}{80.03}{54}
\emline{49.93}{80.03}{55}{47.81}{80.49}{56}
\emline{47.81}{80.49}{57}{45.85}{81.40}{58}
\emline{45.85}{81.40}{59}{44.12}{82.70}{60}
\emline{44.12}{82.70}{61}{42.71}{84.35}{62}
\emline{42.71}{84.35}{63}{41.69}{86.27}{64}
\emline{41.69}{86.27}{65}{41.11}{88.35}{66}
\emline{41.11}{88.35}{67}{40.98}{90.52}{68}
\emline{40.98}{90.52}{69}{41.33}{92.66}{70}
\emline{41.33}{92.66}{71}{42.13}{94.67}{72}
\emline{42.13}{94.67}{73}{43.34}{96.47}{74}
\emline{43.34}{96.47}{75}{44.91}{97.97}{76}
\emline{44.91}{97.97}{77}{46.76}{99.09}{78}
\emline{46.76}{99.09}{79}{48.81}{99.79}{80}
\emline{48.81}{99.79}{81}{51.00}{100.03}{82}
%\end
%\put(60.67,42.33){\makebox(0,0)[cc]
%{Figure 3. Offer and confirmation wave analysis of Afshar setup.}}
%\circle(71.00,66.67){20.35}
\emline{71.00}{76.85}{83}{73.18}{76.61}{84}
\emline{73.18}{76.61}{85}{75.26}{75.91}{86}
\emline{75.26}{75.91}{87}{77.14}{74.78}{88}
\emline{77.14}{74.78}{89}{78.74}{73.28}{90}
\emline{78.74}{73.28}{91}{79.97}{71.47}{92}
\emline{79.97}{71.47}{93}{80.79}{69.43}{94}
\emline{80.79}{69.43}{95}{81.16}{67.27}{96}
\emline{81.16}{67.27}{97}{81.05}{65.08}{98}
\emline{81.05}{65.08}{99}{80.47}{62.96}{100}
\emline{80.47}{62.96}{101}{79.46}{61.01}{102}
\emline{79.46}{61.01}{103}{78.05}{59.33}{104}
\emline{78.05}{59.33}{105}{76.31}{57.99}{106}
\emline{76.31}{57.99}{107}{74.33}{57.06}{108}
\emline{74.33}{57.06}{109}{72.19}{56.57}{110}
\emline{72.19}{56.57}{111}{70.00}{56.54}{112}
\emline{70.00}{56.54}{113}{67.85}{56.99}{114}
\emline{67.85}{56.99}{115}{65.85}{57.89}{116}
\emline{65.85}{57.89}{117}{64.09}{59.20}{118}
\emline{64.09}{59.20}{119}{62.65}{60.86}{120}
\emline{62.65}{60.86}{121}{61.60}{62.78}{122}
\emline{61.60}{62.78}{123}{60.98}{64.89}{124}
\emline{60.98}{64.89}{125}{60.83}{67.08}{126}
\emline{60.83}{67.08}{127}{61.16}{69.24}{128}
\emline{61.16}{69.24}{129}{61.94}{71.29}{130}
\emline{61.94}{71.29}{131}{63.14}{73.13}{132}
\emline{63.14}{73.13}{133}{64.71}{74.66}{134}
\emline{64.71}{74.66}{135}{66.57}{75.83}{136}
\emline{66.57}{75.83}{137}{68.63}{76.57}{138}
\emline{68.63}{76.57}{139}{71.00}{76.85}{140}
%\end
\end{picture}

\normalsize \hskip 6cm $\sigma_1$ \hskip 1.5cm $\sigma_2$
\vskip 1cm

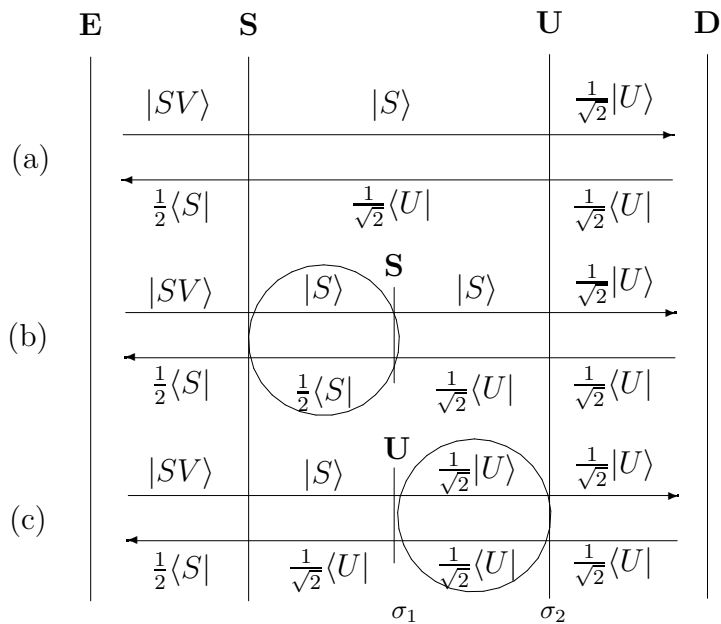
\singlespacing\normalsize \noindent Figure 3. Cramer's diagram modified to
illustrate various aspects of the Afshar setup. (a): The Afshar
experiment without the grid. (b): Inclusion of the grid. (c): A typical
two-slit experiment with ``which-slit'' detectors behind each
slit: downstream from the detectors, the photon is in a determinate
``which-slit'' state.\vskip 1cm 
\onehalfspacing\large
Figure 3(b) represents the inclusion of the wire grid
at $\sigma_1$. We can think of the grid as a confirming,
nondestructive  measurement of the prepared state $|S\rangle$.
Thus we can think of the photon as still being in a superpositon
of slits U and L until it hits the screen at $\sigma_2$
and is forced to ``decide'' which spot it will land in,
U$^\prime$ or L$^\prime$ (the figure only shows the case in which the
outcome happens to be U$^\prime$). 3(c), in which the intervening
measurement is of U, represents what happens in a typical two-slit experiment
with detectors placed behind the slits. We see that downstream from
this measurement, the photon has a determinate ``which-way'' property
(see the circled region), which results in the loss
of interference in subsequent measurements.

What light does this shed on the Afshar experiment? Remember that Afshar
concludes that the ``which-way'' measurement indicates that the
photon ``really'' went through slit U, even though there is an
interference pattern. But the TI picture would disagree, since
the intermediate state is ambiguous, as shown in Figure 3(a)
and to the right of the grid placement in 3(b). Moreover, to the left
of the grid in 3(b), the photon is determinately in a state
of ``went through both slits'': what gets ``disturbed'' by
the grid is not the offer wave, but the confirmation wave! 

The above illustrates the essential agreement between
the TI picture and Unruh's (2004) rejection of Afshar's claim to
have refuted Complementarity. While 3(a) makes it at least
arguable that we could think of the photon as ``fated'' to
choose state U ``out of the starting block,'' since the confirmation
state $\langle U|$ extends back to time $t_0$, in (b) even this
future trace has been obliterated by the intervening S-confirmation
due to the wires at $t_1$. This makes it even less tenable to
attribute the property of ``having gone through slit U'' to the
photon.

Note, however, that the detection of the interference pattern
by the wires at $t_1$ can be taken as confirmation of the
offer wave state $|S\rangle$, and as evidence that the photon is
not in a determinate state with respect to slit location. 
Furthermore, as argued in the previous section,
this means that even when one measures the slit observable $\cal O$
upstream from a screen, and the interference pattern
is lost downstream at the screen, the particle
{\it still went through both slits}--in the sense that the 
offer wave was originally in the superposition state $|S\rangle$. The interference
pattern is lost {\it downstream} from the which-way measurement because,
in the TI picture, what reaches the final screen is the ``particle-like'' offer wave
$|U\rangle$ resulting from the $\cal O$ measurement, as in
Figure 3(c). Moreover, in the time interval between the
which-way measurement and the screen, the particle is in
a determinate slit-basis state with offer and confirmation
wave in agreement. 

Thus TI provides an interesting picture of the
usual delayed-choice experiment (cf. Wheeler 1981), as well. In interpretations
of such experiments, it is often argued that one can influence the past by choosing which
measurement to make in the present.\footnote{\normalsize For example, Wheeler says, ``... in a 
loose sense, we decide what the photon {\it shall have done}
after it has {\it already}\  done it (Wheeler 1983, p. 192).} According to TI, indeed one can---but not
in the thoroughgoing way claimed by Wheeler. That is, according to the 
TI picture a photon heading
toward Earth from a distant star obscured by a large object
acting as a gravitational lens
will {\it always} go ``both ways'' around the object,\footnote{\normalsize Of course,
in this context the term ``photon'' only means ``an excitation
of the electromagnetic field,'' not a particle with a well-localized
trajectory.} regardless
of which measurement we choose to make (which-way or
interference pattern). Our effect on the past is limited
to the confirmation wave that will be sent. If we choose 
to detect an interference pattern, the photon will be in a determinate
state of ``both ways'' throughout the interval; but if
we choose the ``which-way'' measurement, it will be
in an indeterminate state like the photon in Figure 3(a).

Thus, it is consistent to suppose that
a two-slit experiment {\it always} involves a particle
that initially goes through both slits, regardless of what
measurement we choose to perform, because the slits
constitute a preparation of the state S. If, downstream from the 
slits, we do not see an interference pattern, it is because
that offer wave has been altered by an $\cal O$-measurement into an $\cal O$-basis
offer wave, either U or L. But it would not be correct to say that the ``which-way''
measurement indicates that the particle ``actually went
through'' only one or the other slit.

\vskip .5cm
{\bf 5. Conclusion.}
\vskip .5cm

It has been argued that Cramer's Transactional Interpretation provides
a natural and illuminating account of the phenomena reported in
the Afshar experiment. The coexistence of an interference pattern
and a sharp ``slit observable'' $\cal O$ result is explained by way
of offer and confirmation waves, in terms of which the
underlying processes can be visualized. It is pointed out that the
experiment is analogous in key respects to an alleged
paradox described in 1985 by Albert, Aharonov, and D'Amato (AAD)
in the context of a pre- and post-selection experiment,
and that the resolution of the apparent paradox arising
in the Afshar experiment is resolved in the same way. 

Since  Bohr would presumably have no problem with the idea that
we can prepare a particle in a state of ``spin up along $\bf x$'',
confirm that preparation, and then perform a final measurement of
spin along $\bf z$, it is concluded that Afshar's experiment poses no
threat to Complementarity. It is further argued
that the term ``which-way'' measurement is misleading, in that
it tempts one to retrodict from a legitimate measurement of the
slit basis observable that a particle ``really'' went through
one or the other slit, when that may not really be the case.

Finally, it should be emphasized that the arguments presented
in this paper need not be taken as casting doubt on the idea of
measurement in general. In the classical limit, it is certainly
reasonable to infer, for example, that the reading on one's
ammeter reflects what the current was doing one picosecond ago.
It is only when one is dealing with explicit quantum superpositions
(i.e., when the pre- and post-selection measurement observables don't commute)
that one should be cautious about retrodiction, since it is well
known that such retrodictions give rise to paradoxes. On the
other hand, if one wishes to keep the idea of ``which-way'' measurements
and retrodiction even in these cases, then for consistency, one must also accept
retrodiction in the analogous spin case. This, however, puts one in the
dubious position of maintaining that a commonplace spin-measurement experiment
is a violation of Complementarity.

Acknowledgements

I am indebted to John G. Cramer for bringing the Afshar experiment
to my attention, and for valuable correspondence, as well as 
comments from two anonymous reviewers.

\newpage
References.\vskip 1cm

\noindent
Afshar, S. (2004). Sharp complementary wave and particle
behaviors in the same welcher weg experiment. E-print: www.irims.org/quant-ph/030503.\newline
Aharonov, Y., P. Bergmann and J. Lebowitz (1964). Time symmetry
in the quantum process of measurement. {\it Physical Review B, 134,} 1410-1416.\newline
Albert, D., Y. Aharonov, and S. D'Amato (1985). Curious new
statistical prediction of quantum mechanics. {\it Physical Review Letters, 54,} 5-7.\newline
Cramer, John G. (1986). The transactional interpretation
of quantum mechanics. {\it Reviews of Modern Physics, 58,} 647-688. \newline
Unruh, W. (2004). Shahriar Afshar--quantum rebel? E-print: axion.physics.ubc.ca/rebel.html.\newline
Wheeler, J. A. (1981). Delayed-choice experiments and the Bohr-Einstein
dialogue. Originally published in {The American Philosophical
Society and The Royal Society: papers read at a meeting,
June 5, 1980}. Philadelphia: American Philososphical Society.
Reprinted in Wheeler and Zurek (1983, pp. 182-200).\newline
Wheeler, J. A. and W. H. Zurek (1983). {\it Quantum theory
and measurement}. Princeton, N.J.: Princeton University Press.\newline

\end{document}